\documentclass{llncs}

\usepackage{times}
\usepackage[T1]{fontenc} 
\usepackage{url}

\usepackage{latexsym}
\usepackage{xspace}
\usepackage{hyperref}
\usepackage{graphicx}

\usepackage{listings}
\lstnewenvironment{curry}{}{}
\lstnewenvironment{haskell}{}{}
\newcommand{\listline}{\vrule width0pt depth1.75ex}

\newcommand{\code}[1]{\mbox{\small\texttt{#1}}}
\newcommand{\ccode}[1]{``\code{#1}''}
\newcommand{\ol}[1]{\overline{#1}}  

\title{\texorpdfstring
  {A Generic Analysis Server System\\ for Functional Logic Programs}
  {A Generic Analysis Server System for Functional Logic Programs}}
\author{Michael Hanus \and Fabian Reck}
\institute{
Institut f\"ur Informatik, CAU Kiel, D-24098 Kiel, Germany \\
\email{\{mh|fre\}@informatik.uni-kiel.de}
}

\begin{document}
\sloppy
\maketitle

\begin{abstract}
We present a system, called CASS, for the analysis of functional
logic programs.  The system is generic so that various kinds of
analyses (e.g., groundness, non-determinism, demanded arguments) can
be easily integrated.  In order to analyze larger applications
consisting of dozens or hundreds of modules, CASS supports a modular
and incremental analysis of programs.  Moreover, it can be used by
different programming tools, like documentation generators, analysis
environments, program optimizers, as well as Eclipse-based
development environments.  For this purpose, CASS can also be
invoked as a server system to get a language-independent access to
its functionality.  CASS is completely implemented in the functional
logic language Curry as a master/worker architecture to exploit
parallel or distributed execution environments.
\end{abstract}

\section{Introduction}
\label{sec:introduction}

Automated program analyses are useful for various purposes.
For instance, compilers can benefit from their results
to improve the translation of source into target programs.
Analysis information can be helpful to programmers
to reason about the behavior and operational properties
of their programs. Moreover, this information can also be documented
by program documentation tools or interactively shown to developers
in dedicated programming environments.
On the one hand, declarative programming languages provide
interesting opportunities for analyzing programs.
On the other hand, their complex or abstract execution model
demands for good tool support to develop reliable programs.
For example, the detection of type errors in languages with higher-order
features or the detection of mode problems in the use of
Prolog predicates.

This work is related to functional logic languages that
combine the most important
features of functional and logic programming in a single language
(see \cite{AntoyHanus10CACM,Hanus13} for recent surveys).
In particular, these languages provide higher-order functions and demand-driven
evaluation from functional programming together with logic programming features
like non-deterministic search and computing with partial information
(logic variables).
This combination
has led to new design patterns \cite{AntoyHanus02FLOPS,AntoyHanus11WFLP}
and better abstractions for application programming.
Moreover, program implementation and analysis aspects
for functional as well as logic languages can be considered
in a unified framework.
For instance, test cases for functional programs can be generated
by executing functions with logic variables as arguments
\cite{FischerKuchen07}.

Automated program analyses have been already used for functional logic
programming in various situations. For instance,
CurryDoc \cite{Hanus02WFLP} is an automatic documentation tool
for the functional logic language Curry
that analyzes Curry programs to document
various operational aspects, like the non-determinism behavior
or completeness issues.
CurryBrowser \cite{Hanus06WLPE} is an interactive analysis
environment that unifies various program analyses in order
to reason about Curry applications.
KiCS2 \cite{BrasselHanusPeemoellerReck11}, a recent
implementation of Curry that compiles into Haskell,
includes an analyzer to classify higher-order and
deterministic operations in order to support their efficient
implementation which results in highly efficient target programs.
Similar ideas are applied in the implementation
of Mercury \cite{SomogyiHendersonConway96} which uses
mode and determinism information to reorder predicate calls.
Non-determinism information as well as information
about definitely demanded arguments has been used to
improve the efficiency of functional logic programs
with flexible search strategies \cite{Hanus12ICLP}.
A recent Eclipse-based development environment for Curry
\cite{Palkus12} also supports the access to analysis information
during interactive program development.

These different uses of program analyses is the motivation
for the current work.
We present CASS (Curry Analysis Server System)
which is intended to be a central component of
current and future tools for functional logic programs.
CASS provides a generic interface to support the integration
of various program analyses.
Although the current implementation is strongly related to Curry,
CASS can also be used for similar declarative programming languages,
like TOY \cite{Lopez-FraguasSanchez-Hernandez99}.
The analyses are performed on an intermediate format into which 
source programs can be compiled.
CASS supports the analysis of larger applications
by a modular and incremental analysis. 
The analysis results for each module are persistently stored and
recomputed only if it is necessary.
Since CASS is implemented in Curry, it can be directly used
in tools implemented in Curry, like the documentation generator
CurryDoc, the analysis environment CurryBrowser,
or the Curry compiler KiCS2.
CASS can also be invoked as a server system
providing a text-based communication protocol in order to
interact with tools implemented in other languages,
like the Eclipse plug-in for Curry.
CASS is implemented as a master/worker architecture,
i.e., it can distribute the analysis work to different processes
in order to exploit parallel or distributed execution environments.

In the next section, we review some features of Curry.
Section~\ref{sec:anaimpl} shows how various kinds of program analyses
can be implemented and integrated into CASS.
Some uses of CASS are presented in Section~\ref{sec:usage}
before its implementation is sketched in Section~\ref{sec:impl}
and evaluated in Section~\ref{sec:eval}.

\section{Curry and FlatCurry}
\label{sec:flatcurry}

In this section we review some aspects of Curry
that are necessary to understand the functionality and
implementation of our analysis tool.
More details about Curry's computation model and a complete
description of all language features can be found in
\cite{Hanus97POPL,Hanus12Curry}.

Curry is a declarative multi-paradigm language
combining in a seamless way features from functional,
logic, and concurrent programming.
Curry has a Haskell-like syntax\footnote{%
Variables and function names usually
start with lowercase letters and the names of type and data constructors
start with an uppercase letter. The application of $f$
to $e$ is denoted by juxtaposition (``$f~e$'').}
\cite{PeytonJones03Haskell}
extended by the possible inclusion of free (logic)
variables in conditions and right-hand sides of defining rules.
Curry also offers standard features of
functional languages, like polymorphic types, modules, or monadic I/O
which is identical to Haskell's I/O concept \cite{Wadler97}.
Thus, \ccode{IO $\alpha$} denotes the type of an I/O action that returns values
of type $\alpha$.

A \emph{Curry program} consists of the definition of functions
and the data types on which the functions operate.
Functions are defined by conditional equations with constraints in the
conditions.  They are evaluated lazily and can be called with
partially instantiated arguments.
For instance, the following program defines the types of
Boolean values and polymorphic lists
and functions to concatenate lists (infix operator \ccode{++})
and to compute the last element of a list:
\begin{curry}
data Bool   = True | False
data List a = []   | a : List a$\listline$
(++) :: [a] -> [a] -> [a]
[]     ++ ys = ys
(x:xs) ++ ys = x : (xs ++ ys)$\listline$
last xs | _ ++ [x] =:= xs  = x  where x free
\end{curry}
The data type declarations define
\code{True} and \code{False} as the Boolean constants and
\code{[]} (empty list) and \code{:} (non-empty list) as the constructors for
polymorphic lists (\code{a} is a type variable ranging over
all types and the type \ccode{List\,\,a} is usually written as \code{[a]}
for conformity with Haskell).
The (optional) type declaration (\ccode{::}) of the function \code{(++)}
specifies that \code{(++)} takes two lists as input and produces
an output list, where all list elements are of the same
(unspecified) type.\footnote{Curry uses curried function types
where \code{$\alpha$->$\beta$} denotes the type of all functions
mapping elements of type $\alpha$ into elements of type $\beta$.}

The operational semantics of Curry \cite{AlbertHanusHuchOliverVidal05,Hanus97POPL}
is a conservative extension of lazy functional programming (if free variables
do not occur in the program or the initial goal) and (concurrent)
logic programming.
To describe this semantics, compile programs, or implement analyzers and
similar tools, an intermediate representation of Curry programs has been
shown to be useful.
Programs of this intermediate language, called FlatCurry,
contain a single rule for each function where the pattern matching strategy
is represented by case/or expressions.
The basic structure of FlatCurry is defined as follows:\footnote{%
$\ol{o_k}$ denotes a sequence of objects $o_1,\ldots,o_k$.}\\[2ex]
{\small 
$
\begin{array}{@{~~~}lcl@{\hspace*{15ex}}lcl}
P & ::= & D_1 \ldots D_m & e & ::= & x \\
&&& & | & c~e_1\,\ldots\, e_n  \\
D & ::= & f~x_1 \,\ldots\, x_n = e & 
  & | & f~e_1 \,\ldots\, e_n  \\
  & & &
  & | & \mathit{case}~e_0~\mathit{of}~\{\ol{p_k\to e_k}\} \\
p & ::= & c~x_1 \,\ldots\, x_n  &
  & | & \mathit{fcase}~e_0~\mathit{of}~\{\ol{p_k\to e_k}\} \\
  & & &
  & | & e_1~\mathit{or}~e_2 \\
  & & &
  & | & \mathit{let}~\ol{x_k}~\mathit{free~in~} e \\
\end{array}
$}\\[2ex]
A program $P$ consists of a sequence of
function definitions $D$ with pairwise different variables in the left-hand sides.
The right-hand sides are expressions $e$ composed by variables, constructor and
function calls, case expressions, disjunctions,
and introduction of free (unbound) variables.
A case expression has the form
$
\mathit{(f)case}~e~\mathit{of}~
\{c_1~\ol{x_{n_1}} \to e_1,\ldots,c_k~\ol{x_{n_k}} \to e_k\}
$,
where $e$ is an expression, $c_1,\ldots,c_k$ are different 
constructors of the
type of $e$, and $e_1,\ldots, e_k$ are expressions.
The \emph{pattern variables} $\ol{x_{n_i}}$ are local
variables which occur only in the corresponding subexpression $e_i$.
The difference between $\mathit{case}$ and $\mathit{fcase}$ shows up when the
argument $e$ is a free variable:
$\mathit{case}$ suspends (which corresponds to residuation)
whereas $\mathit{fcase}$ nondeterministically binds this variable
to the pattern in a branch of the case expression
(which corresponds to narrowing).

Note that it is possible to translate other functional logic
languages, like TOY \cite{Lopez-FraguasSanchez-Hernandez99},
or even Haskell into this intermediate format.
Since our analysis tool is solely based on FlatCurry,
it can also be used for other source languages provided that
there is a translator from such languages into FlatCurry.

Mature implementations of Curry, like PAKCS \cite{Hanus13PAKCS}
or KiCS2 \cite{BrasselHanusPeemoellerReck11}, provide support
for meta-programming by a library containing data types
for representing FlatCurry programs
and an I/O action for reading a module and translating
its contents into the corresponding data term.
For instance, a module of a Curry program
is represented as an expression of type
\begin{curry}
data Prog = Prog String [String] [TypeDecl] [FuncDecl] [OpDecl]
\end{curry}
where the arguments of the data constructor \code{Prog}
are the module name, the names of all imported modules,
the list of all type, function, and infix operator declarations.
Furthermore, a function declaration is represented as
\begin{curry}
data FuncDecl = Func QName Int Visibility TypeExpr Rule
\end{curry}
where the arguments are the qualified name
(of type \code{QName}, i.e., a pair of module and function name),
arity, visibility (\code{Public} or \code{Private}), type, and rule
(of the form \ccode{Rule $\mathit{arguments}$ $\mathit{expr}$})
of the function.
Finally, the data type for expressions just reflects its
formal definition:\footnote{We present a slightly simplified version
of the actual type definitions.}
\begin{curry}
data Expr = Var Int
          | Lit Literal
          | Comb CombType QName [Expr]
          | Case CaseType Expr [(Pattern,Expr)]
          | Or Expr Expr
          | Free [Int] Expr$\listline$
data CombType = FuncCall | ConsCall
\end{curry}
Thus, variables are numbered, literals (like numbers or characters)
are distinguished from combinations (\code{Comb}) which
have a first argument to distinguish constructor applications
and applications of defined functions.
The remaining data type declarations for representing Curry programs
are similar but we omit them for brevity.

\section{Implementing Program Analyses}
\label{sec:anaimpl}

Basically, a program analysis can be considered as a mapping
that associates a program element with information about some 
aspect of its semantics.
Since most interesting semantical aspects are not computable,
they are approximated by some abstract domain
where each abstract value describes some set of concrete values
\cite{CousotCousot77}.
For instance, an ``overlapping rules'' analysis determines
whether a function is defined by
a set of overlapping rules, which means that
some ground call to this function can be reduced in
more than one way.
An example for an operation that is defined by overlapping
rules is the ``choice'' operation
\begin{curry}
x ? y = x
x ? y = y
\end{curry}
For this analysis one can use \code{Bool} as the abstract domain
so that the abstract value \code{False} is interpreted
as ``defined by non-overlapping rules'' and \code{True}
is interpreted as ``defined by overlapping rules''.
Hence, the ``overlapping rules'' analysis has the type
\begin{curry}
FuncDecl -> Bool
\end{curry}
which means that we associate a \code{Bool} value to each function definition.
Based on the data type definitions sketched in
Section~\ref{sec:flatcurry} and some standard functions,
such an analysis can be defined by looking for occurrences of \code{Or}
in the defining expression as follows:
\begin{curry}
isOverlapping :: FuncDecl -> Bool
isOverlapping (Func _ _ _ _ (Rule _ e))   = orInExpr e$\listline$
orInExpr :: Expr -> Bool
orInExpr (Var _)       = False
orInExpr (Lit _)       = False
orInExpr (Comb _ _ es) = any orInExpr es
orInExpr (Case _ e bs) = orInExpr e || any (orInExpr . snd) bs
orInExpr (Or _ _)      = True
orInExpr (Free _ e)    = orInExpr e
\end{curry}
Many interesting aspects require a more sophisticated analysis
where dependencies are taken into account.
For instance, consider a ``non-determinism'' analysis
with the abstract domain
\begin{curry}
data DetDom = Det | NonDet
\end{curry}
Here, \code{Det} is interpreted as ``the operation evaluates
in a deterministic manner on ground arguments.''
However, \code{NonDet} is interpreted as ``the operation \emph{might}
evaluate in different ways for given ground arguments.''
The apparent imprecision is due to the approximation of the analysis.
For instance, if the function \code{f} is defined by overlapping rules
and the function \code{g} might call \code{f}, then \code{g}
is judged as non-deterministic.
In order to take into account such dependencies,
the non-determinism analysis requires to examine the current
function as well as all directly or indirectly called functions
for overlapping rules.
Due to recursive function definitions, this analysis cannot be done
in one shot but requires a fixpoint computation.
In order to make things simple for the analysis developer,
CASS supports such fixpoint computations and requires from the
developer only the implementation of an operation of type
\begin{curry}
FuncDecl -> [(QName,a)] -> a
\end{curry}
where \ccode{a} denotes the type of abstract values.
The second argument of type \code{[(QName,a)]}
represents the currently known analysis values
for the functions \emph{directly} used in this function declaration.
Hence, the non-determinism analysis can be implemented
as follows:
\begin{curry}
nondetFunc :: FuncDecl -> [(QName,DetDom)] -> DetDom
nondetFunc (Func f _ _ _ (Rule _ e)) calledFuncs =
  if orInExpr e || freeVarInExpr e ||
     any (==NonDet) (map snd calledFuncs)
  then NonDet
  else Det
\end{curry}
Thus, it computes the abstract value \code{NonDet}
if the function itself is defined by overlapping rules or
contains free variables that might cause non-deterministic guessing
(we omit the definition of \code{freeVarInExpr} since it is quite
similar to \code{orInExpr}), or
if it depends on some non-deterministic function.
The actual analysis is performed by defining some start value
for all functions (the ``bottom'' value of the abstract domain,
here: \code{Det}) and performing a fixpoint computation for
the abstract values of these functions.
CASS uses a working list approach as default
but also supports other methods to compute fixpoints.
The termination of the fixpoint computation can be ensured
by standard assumptions in abstract interpretation
\cite{CousotCousot77}, e.g., by choosing a finite abstract domain
and monotonic operations, or by widening operations.

To support the inclusion of different analyses in CASS,
there are an abstract type
\ccode{Analysis a}
denoting a program analysis with abstract domain \ccode{a}
and several constructor operations for various kinds of analyses.
Each analysis has a name provided as a first argument
to these constructors. The name is used to store the
analysis information persistently and to pass specific analysis tasks
to workers (see below for more details).
For instance, a simple function analysis which depends only on a
given function definition can be created by
\begin{curry}
funcAnalysis :: String -> (FuncDecl -> a) -> Analysis a
\end{curry}
where the analysis name and analysis function are provided as arguments.
Thus, the overlapping analysis can be specified as
\begin{curry}
overlapAnalysis :: Analysis Bool
overlapAnalysis = funcAnalysis "Overlapping" isOverlapping
\end{curry}
A function analysis with dependencies can be constructed by
\begin{curry}
dependencyFuncAnalysis ::
  String -> a -> (FuncDecl -> [(QName,a)] -> a) -> Analysis a
\end{curry}
Here, the second argument specifies the start value of the fixpoint analysis,
i.e., the bottom element of the abstract domain.
Thus, the complete non-determinism analysis can be specified as
\begin{curry}
nondetAnalysis :: Analysis DetDom
nondetAnalysis = dependencyFuncAnalysis "Deterministic" Det nondetFunc
\end{curry}
It should be noted that this definition is sufficient to execute the analysis
with CASS since the analysis system takes care of computing fixpoints,
calling the analysis functions with appropriate values,
analyzing imported modules, etc.
Thus, the programmer can concentrate on implementing
the logic of the analysis and is freed from many tedious implementation
details.

Sometimes one is also interested in analyzing information about
data types rather than functions.
For instance, the Curry implementation
KiCS2 \cite{BrasselHanusPeemoellerReck11} has an optimization
for higher-order deterministic operations.
This optimization requires some information about the
higher-order status of data types, i.e., whether a term of some type
might contain functional values.
CASS supports such analyses by appropriate analysis constructors.
A simple type analysis which depends only on a given
type declaration can be specified by
\begin{curry}
typeAnalysis :: String -> (TypeDecl -> a) -> Analysis a
\end{curry}
A more complex type analysis depending also on information
about the types used in the type declaration can be specified by
\begin{curry}
dependencyTypeAnalysis ::
  String -> a -> (TypeDecl -> [(QName,a)] -> a) -> Analysis a
\end{curry}
Similarly to a function analysis, the second argument is the start value
of the fixpoint analysis and the third argument computes
the abstract information about the type names used in the type declaration.

The remaining entities in a Curry program that can be analyzed
are data constructors. Since their definition only contains the
argument types, it may seem uninteresting to provide a useful
analysis for them. However, sometimes it is interesting
to analyze their context so that there is a analysis constructor
of type
\begin{curry}
constructorAnalysis :: String -> (ConsDecl -> TypeDecl -> a)
                    -> Analysis a
\end{curry}
Thus, the analysis operation of type \code{(ConsDecl -> TypeDecl -> a)}
gets for each constructor declaration the corresponding type declaration.
This information could be used to compute the sibling constructors, e.g.,
the sibling for the constructor \code{True} is \code{False}.
The information about sibling constructors is useful to check
whether a function is completely defined, i.e., contains
a case distinction for all possible patterns.
For instance, the operation (in FlatCurry notation)
\begin{curry}
not x = case x of True  -> False
                  False -> True
\end{curry}
is completely defined whereas
\begin{curry}
cond x y = case x of True -> y
\end{curry}
is incompletely defined since it fails on \code{False} as the first
argument.
To check this property, information about sibling constructors
is obviously useful.
But how can we provide this information in an analysis for functions?

For this and similar purposes, CASS supports the
combination of different analyses.
Thus, an analysis developer can define an analysis
that is based on information computed by another analysis.
To make analysis combination possible,
there is an abstract type \ccode{ProgInfo a}
to represent the analysis information of type \ccode{a}
for a given module and its imports.
In order to look up analysis information about some entity, there is
an operation
\begin{curry}
lookupProgInfo:: QName -> ProgInfo a -> Maybe a
\end{curry}
One can use the analysis constructor
\begin{curry}
combinedFuncAnalysis :: String -> Analysis b
                  -> (ProgInfo  b -> FuncDecl -> a) -> Analysis a
\end{curry}
to implement a function analysis depending on some other analysis.
The second argument is some base analysis computing abstract values
of type \ccode{b} and the analysis function gets, in contrast to a simple
function analysis, the analysis information computed by this base
analysis.

For instance, if the sibling constructor analysis is defined as
\begin{curry}
siblingCons :: Analysis [QName]
siblingCons = constructorAnalysis $\ldots$
\end{curry}
then the pattern completeness analysis might be defined by
\begin{curry}
patCompAnalysis :: Analysis Bool
patCompAnalysis =
  combinedFuncAnalysis "PatComplete" siblingCons isPatComplete$\listline$
isPatComplete :: ProgInfo [QName] -> FuncDecl -> Bool
isPatComplete siblinginfo fundecl = $\ldots$
\end{curry}
Similarly, other kinds of analyses can be also combined with some base
analysis by using the following analysis constructors:
\begin{curry}
combinedTypeAnalysis :: String -> Analysis b
                  -> (ProgInfo  b -> TypeDecl -> a) -> Analysis a
combinedDependencyFuncAnalysis :: String -> Analysis b -> a
   -> (ProgInfo b -> FuncDecl -> [(QName,a)] -> a) -> Analysis a
combinedDependencyTypeAnalysis :: String -> Analysis b -> a
   -> (ProgInfo b -> TypeDecl -> [(QName,a)] -> a) -> Analysis a
\end{curry}
For instance, an analysis for checking whether a function
is totally defined, i.e., always reducible for all ground arguments,
can be based on the pattern completeness analysis.
It is a dependency analysis so that it can be defined as follows
(in this case, \code{True} is the bottom element since the
abstract value \code{False} denotes ``might not be totally defined''):
\begin{curry}
totalAnalysis :: Analysis Bool
totalAnalysis =
  combinedDependencyFuncAnalysis "Total" patCompAnalysis True isTotal$\listline$
isTotal :: ProgInfo Bool -> FuncDecl -> [(QName,Bool)] -> Bool
isTotal pcinfo fdecl calledfuncs =
  (maybe False id (lookupProgInfo (funcName fdecl) pcinfo))
  && all snd calledfuncs
\end{curry}
Hence, a function is totally defined if it is pattern complete
and depends only on totally defined functions.

Further examples of combined analyses are the higher-order function
analysis used in KiCS2 (see above) where the higher-order status
of a function depends on the higher-order status of its argument types,
and the non-determinism analysis of \cite{BrasselHanus05}
where non-determinism effects are analyzed based on groundness
information.

In order to integrate some implemented analysis in CASS,
one has to register it. In principle, this could be done dynamically,
but currently only a static registration is supported.
For the registration, the implementation of CASS contains a constant
\begin{curry}
registeredAnalysis :: [RegisteredAnalysis]
\end{curry}
keeping the information about all available analyses.
To register a new analysis, it has to be added to this list of
registered analyses and CASS has to be recompiled.
Each registered analysis must provide a ``show'' function
to map abstract values into strings to be shown to the user.\footnote{%
Alternative visualizations of analysis information, e.g., as graphs,
are planned for the future.}
This allows for some flexibility in the presentation of the analysis information.
For instance, showing the results of the totally definedness analysis
can be implemented as follows:
\begin{curry}
showTotal :: Bool -> String
showTotal True  = "totally defined"
showTotal False = "possibly partially defined"
\end{curry}
An analysis can be registered with the auxiliary operation
\begin{curry}
cassAnalysis :: Analysis a -> (a -> String) -> RegisteredAnalysis
\end{curry}
For instance, we can register our analyses presented in this section
by the definition
\begin{curry}
registeredAnalysis = [cassAnalysis overlapAnalysis  showOverlap
                     ,cassAnalysis nondetAnalysis   showDet
                     ,cassAnalysis siblingCons      showSibling
                     ,cassAnalysis patCompAnalysis  showComplete
                     ,cassAnalysis totalAnalysis    showTotal    ]
\end{curry}
in the CASS implementation.
After compiling CASS, they are immediately available
as shown in the next section.

\section{Using the Analysis System}
\label{sec:usage}

As mentioned above, a program analysis
is useful for various purposes, e.g., the implementation
and transformation of programs, tool and documentation support
for programmers, etc.
Therefore, the results computed by some analysis registered
in CASS can be accessed in various ways.
Currently, there are three methods for this purpose:
\begin{description}
\item[Batch mode:]
CASS is started with a module and analysis name.
Then this analysis is applied to the module and the
results are printed (using the analysis-specific show function, see above).
\item[API mode:]
If the analysis information should be used in an application
implemented in Curry, the application program could use the CASS interface
operations to start an analysis and use the computed results
for further processing.
\item[Server mode:]
If the analysis results should be used in an application
implemented in some language that does not have a direct interface to Curry,
one can start CASS in a server mode. In this case, one can connect to
CASS via some socket using a communication protocol that is specified
in the documentation of CASS.
\end{description}
\begin{figure*}[t]
\begin{center}
\includegraphics[width=\textwidth]{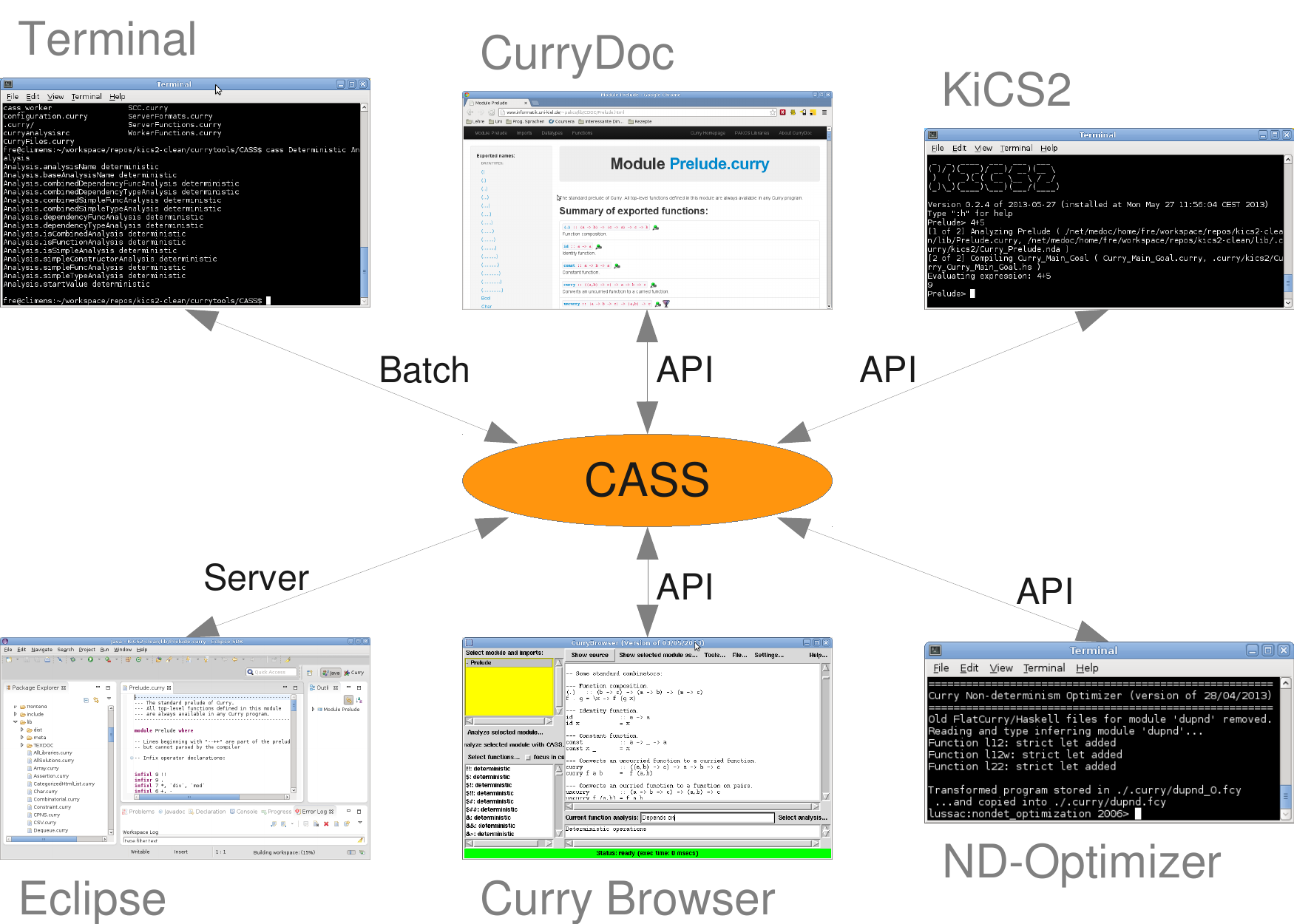}
\end{center}
\caption{Using CASS in different contexts\label{fig:cass-usage}}
\end{figure*}
Figure~\ref{fig:cass-usage} shows some uses of CASS
which are discussed in the following.
The use of CASS in batch mode is obvious.
This mode is useful to get a quick access to analysis information
so that one can experiment with different abstractions,
fixpoint computations, etc.

If one wants to access CASS inside an application implemented in
Curry, one can use some interface operation of CASS.
For instance, CASS provides an operation
\begin{curry}
analyzeGeneric :: Analysis a -> String
              -> IO (Either (ProgInfo a) String)
\end{curry}
to apply an analysis (first argument) to some module (whose name
is given in the second argument).
The result is either the analysis information computed for this module
or an error message in case of some execution error.
This access to CASS is used in the documentation generator CurryDoc
\cite{Hanus02WFLP} to describe some operational aspects of functions
(e.g., pattern completeness, non-determinism, solution completeness),
the Curry compiler KiCS2 \cite{BrasselHanusPeemoellerReck11}
to get information about the determinism and higher-order status of
functions, and the non-determinism optimizer described in \cite{Hanus12ICLP}
to obtain information about demanded arguments and non-deterministic
functions. Furthermore, there is also a similar operation
\begin{curry}
analyzeModule :: String -> String
              -> IO (Either (ProgInfo String) String)
\end{curry}
which takes an analysis name and a module name as arguments and
yields the textual representation of the computed analysis results.
This is used in the CurryBrowser \cite{Hanus06WLPE}
which allows the user to browse through the modules of
a Curry application and apply and visualize various analyses
for each module or function. Beyond some specific analyses
like dependency graphs, all function analyses registered in CASS
are automatically available in the CurryBrowser.

The server mode of CASS is used in a recently developed
Eclipse plug-in for Curry \cite{Palkus12} which also supports
the visualization of analysis results inside Eclipse.
Since this plug-in is implemented in a Java-based framework,
the access to CASS is implemented via a textual protocol
over a socket connection.
This protocol has a command \code{GetAnalysis} to query
the names of all available analyses.
This command is used to initialize the
analysis selection menus in the Eclipse plug-in.
Furthermore, there are commands to analyze a complete module
or individual entities inside a module.
The analysis results are returned as plain strings or in
XML format.
Currently, we are working on more options to visualize
analysis information in the Eclipse plug-in rather than strings,
e.g., term or graph visualizations.

\section{Implementation}
\label{sec:impl}

As mentioned above, CASS is implemented in Curry using the
features for meta-programming as sketched in
Section~\ref{sec:flatcurry}.
Since the analysis programmer only provides operations
to analyze a function, type, or data constructor,
as shown in Section~\ref{sec:anaimpl},
the main task of CASS is to supply these operations with
the appropriate parameters in order to compute the analysis
results.

CASS is intended to analyze larger applications consisting of
many modules. Thus, a simple implementation by concatenating
all modules into one large program to be analyzed would not
be efficient enough.
Hence, CASS performs a separate analysis of each module
by the following steps:
\begin{enumerate}
\item The imported modules are analyzed.
\item The analysis information of the interface of the imported modules
are loaded.
\item The module is analyzed. If the analysis is a dependency analysis,
they are evaluated by a fixpoint computation where the specified start
value is used as initial values for the locally defined (i.e., non-imported)
entities.
\end{enumerate}
Obviously, this scheme can be simplified in case of a simple
analysis without dependencies, since such an analysis does not require
the imported entities.
For a combined analysis, the base analysis is performed
before the main analysis is executed.

It should be noted that the separate analysis of each module
allows only a bottom-up but not a top-down analysis starting with the initial
goal. A bottom-up analysis is sufficient for interactive systems where the 
initial goal is not known at analysis time.
Nevertheless, it is sometimes possible to express ``top-down oriented''
analyses, like a groundness analysis, in a bottom-up manner
by choosing appropriate abstract domains, as shown in
\cite{BrasselHanus05} where a type and effect system is used
to analyze groundness and non-determinism information.

In order to speed up the complete analysis process,
CASS implements a couple of improvements to this general analysis
process sketched above.
First, the analysis information for each module is persistently stored.
Hence, before a module is analyzed, it is checked whether
there already exists a storage with the analysis information
of this module and whether the time stamp of this information is newer
than the source program with all its direct or indirect imports.
If the storage is found and is still valid, the stored information is used.
Otherwise, the information is computed as described above
and then persistently stored.
This has the advantage that, if only the main module has changed and needs to be
re-analyzed, the analysis time of a large application is still small.

To exploit multi-core or distributed execution environments,
the implementation of CASS is designed as a master/worker architecture
where a master process coordinates all analysis activities
and each worker is responsible to analyze a single module.
Thus, when CASS is requested to analyze some module,
the master process computes all import dependencies
together with a topological order of all dependencies.
Therefore, the standard prelude module (without import dependencies)
is the first module to be analyzed and the main module is the last one.
Then the master process iterates on the following steps
until all modules are analyzed:
\begin{itemize}
\item If there is a free worker and all imports of the first module
are already analyzed, pass the first module to the free worker
and delete it from the list of modules.
\item If the first module contains imports that are not yet analyzed,
wait for the termination of an analysis task of a worker.
\item If a worker has finished the analysis of a module,
mark all occurrences of this module as ``analyzed.''
\end{itemize}
Since contemporary Curry implementations do not support
thread creation, the workers are implemented as processes
that are started at the beginning and terminated at the end of the
entire execution. The number of workers can be defined by some
system parameter.

The current distribution of CASS\footnote{CASS is part of the distributions
of the Curry systems KiCS2 \cite{BrasselHanusPeemoellerReck11}
and PAKCS \cite{Hanus13PAKCS}.} contains fourteen program analyses,
including the analyses discussed in Section~\ref{sec:anaimpl}.
Further analyses include
a ``solution completeness'' analysis (which checks whether a function
might suspend due to residuation),
a ``right-linearity'' analysis (used to improve the implementation
of functional patterns \cite{AntoyHanus05LOPSTR}),
an analysis of demanded arguments (used to optimize non-deterministic
computations \cite{Hanus12ICLP}),
or a combined groundness/non-determinism analysis based on a
type and effect system \cite{BrasselHanus05}.
New kinds of analyses can easily be added, since,
as shown in Section~\ref{sec:anaimpl},
the infrastructure provided by CASS simplifies their definition and
integration.

\section{Practical Evaluation}
\label{sec:eval}

We have already discussed some practical applications of CASS
in Section~\ref{sec:usage}.
These applications demonstrate that the current implementation
with a module-wise analysis,
storing analysis information persistently, and
incremental re-analysis is good enough to use CASS in practice.
In order to get some ideas about the efficiency of the
current implementation, we made some benchmarks
and report their results in this section.
Since all analyses contained in CASS have been developed and
described elsewhere (see the references above),
we do not evaluate their precision but only their execution efficiency.

CASS is intended to analyze larger systems.
Thus, we omit the data for analyzing single modules
but present the analysis times for four different
Curry applications:
the interactive environment (read/eval/print loop)
of KiCS2,
the analysis system presented in this paper,
the interactive analysis environment CurryBrowser \cite{Hanus06WLPE},
and the module database, a web application generated from
an entity/relationship model with the web framework
Spicey \cite{HanusKoschnicke10PADL}.
In order to get an impression of the size of each application,
the number of modules (including imported system modules)
is shown for each application.
Typically, most modules contain between 100-300 lines of code,
where the largest one has more than 900 lines of code.

Table~\ref{table:benchmarks} contains the elapsed time (in seconds)
needed to analyze
these applications for different numbers of workers.
We ran two kinds of fixpoint analysis: an analysis of demanded arguments
\cite{Hanus12ICLP} and a groundness analysis \cite{BrasselHanus05}.
Each analysis has always been started from scratch, i.e., all persistently
stored information were deleted at the beginning, except for
the last row which shows the times to re-analyze the application
where only the main module has been changed.
In this case, the actual analysis time is quite small
but most of the total time is spent to check all module dependencies
for possible updates.
The benchmarks were executed on a Linux machine running
Ubuntu 12.04 with an Intel Core i5 (2.53GHz) processor
where CASS was compiled with KiCS2 (Version 0.2.4).

\begin{table}[t]
\begin{center}
\begin{tabular}{|c|*{8}{r|}}
\hline
Application: & \multicolumn{2}{c|}{KiCS2 REPL} & \multicolumn{2}{c|}{CASS} & \multicolumn{2}{c|}{CurryBrowser} & \multicolumn{2}{c|}{ModuleDB} \\
\hline
Modules:     & \multicolumn{2}{c|}{ 32  } & \multicolumn{2}{c|}{ 46 } & \multicolumn{2}{c|}{  71   } & \multicolumn{2}{c|}{  85    } \\
\hline
Analysis:  & Demand & Ground & Demand & Ground & Demand & Ground & Demand & Ground \\
\hline
1 worker:  & 8.09 & 8.25 & 10.25 & 10.30 & 19.53 & 19.36 & 27.97 & 28.15 \\
\hline
2 workers: & 5.75 & 5.82 &  6.87 &  7.48 & 12.33 & 12.49 & 18.32 & 18.56 \\
\hline
4 workers: & 5.41 & 5.47 &  6.17 &  6.47 & 10.20 & 10.38 & 16.98 & 17.15 \\
\hline
Re-analyze:& 1.40 & 1.38 &  1.26 &  1.26 &  2.01 &  1.99 &  2.34 &  2.34 \\
\hline
\end{tabular}
\end{center}
\caption{Using CASS in different contexts\label{table:benchmarks}}
\end{table}

The speedup related to the number of workers is not optimal.
This might be due to the fact that the dependencies between the modules
are complex so that there are not many opportunities for an independent
analysis of modules, i.e., workers might have to wait for the
termination of the analysis of modules which are imported by many
other modules.
Nevertheless, the approach shows that there is a potential to exploit
the computing power offered by modern computers.
Furthermore, the absolute run times are acceptable.
It should also be noted that, during system development, the times are lower
due to the persistent storing of analysis results.

\section{Conclusions}
\label{sec:conclusions}

In this paper we presented CASS, a tool to analyze functional logic programs.
CASS supports various kinds of program analyses
by a general notion of analysis functions that map
program entities into analysis information.
In order to implement an analysis that also depends on information
about other entities used in a definition, CASS supports
``dependency analyses'' that require a fixpoint computation
to yield the final analysis information.
Moreover, different analyses can be combined so that
one can define an analysis that is based on the results of another analysis.
Using these different constructions,
the analysis developer can concentrate on defining the logic
of the analysis and is freed from the details to invoke
the analysis on modules and complete application systems.
To analyze larger applications efficiently,
CASS performs a modular and incremental analysis
where already computed analysis information is persistently stored.
Thus, CASS does not support top-down or goal-oriented analyses
but only bottom-up analyses which is acceptable for large applications
or interactive systems with unknown initial goals.
The implementation of CASS supports different modes of use
(batch, API, server) so that the registered analyses
can be accessed by various systems, like compilers, program optimizers,
documentation generators, or programming environments.
Currently, CASS produces output in textual form.
The support for other kinds of visualizations is a topic for future work.

The analysis of programs is an important topic for all kinds
of languages so that there is a vast body of literature.
Most of such works is related to the development and application
of various analysis methods (where some of them related
to functional logic programs have already been discussed
in this paper), but there are less works on the development
or implementation of program analyzers.
An example of such an approach, that is in some aspects
similar to our work, is Hoopl \cite{RamseyDiasPeytonJones10}.
Hoopl is a framework for data flow analysis and transformation.
As our framework does, Hoopl eases the definition of
analyses by offering high-level abstractions and
releases the user from tasks like writing fixpoint computations.
In contrast to our work, Hoopl works on a generic representation
of data flow graphs, whereas CASS performs 
incremental, module-wise analyses on an already existing
representation of functional logic programs. 
Another related system is Ciao \cite{HermenegildoEtAl12},
a logic programming system with advanced program analysis features
to optimize and verify logic programs.
CASS has similar goals but supports strongly typed analysis constructors
to make the analysis construction reliable.

There are only a few approaches or tools directly related
to the analysis of combined functional logic programs, as already
discussed in this paper.
The examples in this paper show that this combination is valuable
since analysis aspects of pure functional and pure logic languages
can be treated in this combined framework, like demand and higher-order
aspects from functional programming and groundness and determinism
aspects from logic programming.
An early system in this direction is CIDER \cite{HanusKoj01WFLP}.
CIDER supports the analysis of single Curry modules
together with some graphical tracing facilities.
A successor of CIDER is CurryBrowser \cite{Hanus06WLPE},
already mentioned above, which supports the analysis
and browsing of larger applications.
CASS can be considered as a more efficient and more general implementation
of the analysis component of CurryBrowser.

For future work, we will add further analyses in CASS with more advanced
abstract domains. Since this might lead to analyses with substantial
run times,
the use of parallel architectures might be more relevant.
Thus, it would be also interesting to develop advanced methods
to analyze module dependencies in order to obtain a better distribution
of analysis tasks between the workers.

\paragraph{Acknowledgements.}
The authors are grateful to Heiko Hoffmann for his contribution
to an initial version of the analysis system and Sandra Dylus
for her suggestions to improve the paper.


\end{document}